# Study of Ni-Zn battery by coupling Strain Gauge measurements and Acoustic Emission

F. Degret[1,2], S. Lespinats[1], N. Guillet[1], M. Alias[1], P.-X. Thivel[2]
[1]Univ. Grenoble Alpes, CEA, LITEN, DTS, F-38000 Grenoble
[2]Univ. Grenoble Alpes, CNRS, LEPMI, F-38000 Grenoble, France

**Abstract :**
Two different non-destructive techniques of characterization were coupled to study the mechanical behavior of materials inside a battery during operation: measurement of the deformation of a battery casing by strain gauge and acoustic emission due to the release of mechanical stress inside the battery materials. Experiments were conducted on a commercial Ni-Zn AA-type battery and we particularly considered the phenomena that occurs during overcharge. Beyond the example of this study on a Ni-Zn battery, these two techniques offer complementary information that can be very useful for the monitoring and management of different type of batteries.

**Graphical abstract**

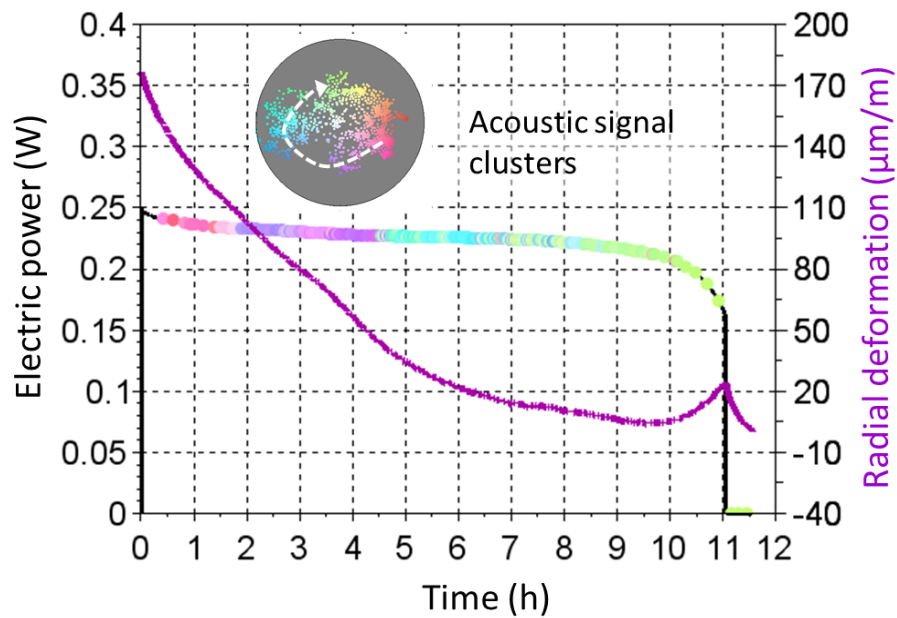

## 1. Introduction

Performance and safety monitoring of the batteries is of major importance for most of applications. However, present methods of battery monitoring embedded in the Battery Management Systems (BMS) are usually only based on temperature and electrochemical parameters measurements (cell voltage, applied current, coulombic counting…). They do not provide relevant enough information to detect the mild signs of improper operation that could lead to irreversible degradation of the performance and even safety issues.

Development of BMS that allows a better understanding of the reactions occurring inside the batteries is required to increase the battery cycle life and performance, as well as their safety. Indeed, an accurate monitoring of the battery could allow to adapt constantly the operation limits of the battery (voltage range, max charge and discharge current) to the specific conditions of operation, instead of using arbitrarily fixed values, conservatively defined for batteries at the beginning of their life. Such a management of the battery requires collecting relevant information related to the battery behavior during operation.

To address this fact, we considered the use of two complementary characterization techniques to complete the usual electrochemical monitoring of batteries: strain gauges measurements to record the battery cell deformation and acoustic emission (AE) to detect transient elastic waves propagating throughout the battery subsequently to the release of internal mechanical stresses in the battery's materials.

The strain gauge signal is given as an electrical resistance variations involved by the deformations of a sensor placed at the surface of the sample considered. A strain gauge sensor is constituted of a small electrically resistive circuit made of several pieces, bonded to the sample studied. Strain gauges are often assembled following 3 distinct orientations, like the rectangular rosette gauge, the tee rosette gauge or the delta rosette gauge, which allows to position freely the rosette without knowing the principal deformation axis. This technique have already been used to study electrochemical systems [1,2] and the batteries deformation measurement could give relevant additional information about the mechanisms occurring inside the batteries during operation [3,4] as well as the possibility to estimate the state of charge and state of health [5,6].

Acoustic emission technique is a non-destructive technique (NDT) used to detect the transient elastic waves spontaneously generated by the release of mechanical stress accumulated in materials and at interfaces. The elastic waves, which propagate into the material provides acoustic signals, allowing the detection of events in real time during the test [7]. This technique is widely used in the industry either for process monitoring, material characterization or damage assessment. The variety of deformation and damage mechanisms that can be sources of AE is numerous. Among them, AE has already been used to study or monitor electrochemical phenomenon as corrosion [8,9] or electrolysis [10,11]. Concerning batteries, phenomenon like lithium insertion, passivation film formation or pulverization of electrodes have already been the subject of many studies [7,12–17].

The main objective of this work is to evaluate the ability and limitations of these two non-electrochemical, non-invasive and *operando* techniques to detect subtle materials and interfaces changes during operation (e.g. due to volume change, cracks, gas evolution, etc…).

This evaluation was conducted on commercial nickel – zinc (Ni-Zn) batteries. This battery is an alkaline rechargeable system developed at the end of the 19[th] century. The invention was patented by Titus von Michalowski in 1899 [18,19], and then, by Thomas Edison in 1901 [20]. The advantages of this system are related to the high cell voltage (theoretical open-circuit voltage of the electrochemical couple is 1.73 V), compared to other nickel-based batteries (1.4 V for Ni-Fe, 1.35 V for Ni-MH), the high theoretical specific energy (334 Wh.kg$^{-1}$, close to the value of theoretical value of Li-ion: 400 Wh.kg$^{-1}$, whereas only 70 to 80 Wh.kg$^{-1}$ is reached for the best Ni-Zn batteries currently commercially available), the use of abundant materials with low environmental impact and are easily recyclable [21,22]. The technology was for a long time limited by the low cycle life of the battery due to the complexity of the solubility of zinc in the alkaline electrolyte [23–27]. Improvements made since the 1970's to stabilize the zinc electrode enhanced the cycle life and increased interest of this battery as power source for a number of applications including stationary storage, vehicles such as electric bicycles and scooters, wireless equipment... [22,28–31]

Nickel – zinc battery uses the nickel oxyhydroxide / nickel hydroxide (NiOOH / Ni(OH)$_2$) electrochemical couple as positive electrode and zinc / zinc hydroxide (Zn / Zn(OH)$_2$) as negative electrode.

Discharge:

Positive electrode: $2\ NiOOH + 2\ H_2O + 2\ e^- \rightarrow 2\ Ni(OH)_2 + 2\ OH^-$ 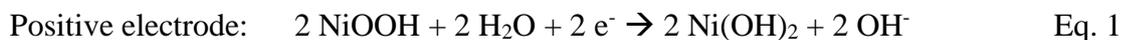 Eq. 1

Negative electrode: $Zn + 2\ OH^- \rightarrow Zn(OH)_2 + 2\ e^-$ 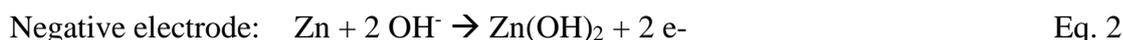 Eq. 2

Overall reaction: $2\ NiOOH + 2\ H_2O + Zn \rightarrow 2\ Ni(OH)_2 + Zn(OH)_2$ 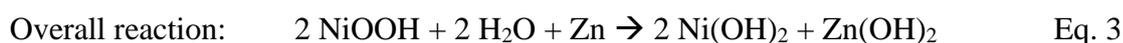 Eq. 3

The nickel electrode is roughly the same as the electrode used in other technologies like nickel – cadmium, nickel – iron, nickel - metal hydride [29]. It is made of a highly porous structure of nickel metal that also acts as current collector. The zinc negative electrode is generally prepared in the discharged state, depositing a paste of zinc oxide onto a mechanical support that is also used as negative current collector (e.g. copper foam, tin-plated copper expanded foil [30–32]). The electrodes are place in a alkaline aqueous electrolyte (typically 25 *wt.*% KOH [28,32]) and a separator is placed in between the electrodes [30,31].

Life cycle limitation is attributed to several issues, among them are the physical change leading to mechanical stress of the electrodes during cycling and overcharge/overdischarge:

- During discharge, zinc is oxidized and can form complex zincate ions (e.g. Zn(OH)$_4^{2-}$) that are partially soluble in alkaline solution. They naturally diffuse throughout the battery and precipitate as zinc oxide when the solubility limit is reached. The solubility of the zincate highly depends on the temperature and the pH that can locally change a lot as two molecule of H₂O are consumed each time a zinc atom is oxidized. Precipitation of the oxide may cause clogging of the porous structure and form of an insulating zinc oxide layer at the surface of the negative electrode. During the next charge, zinc is not uniformly deposited on the electrode and the phenomenon increases with cycles, eventually causing electrode expansion and contraction and even dendritic growth.

- Another problem is the hydrogen and oxygen evolution through parasitic reactions, when the battery is overcharged or over discharged [33–36]. During overcharge, oxygen evolution reaction (OER) occurs at the nickel electrode and hydrogen evolution reaction (HER) can be observed at the zinc electrode. If the battery is over-discharged, hydrogen can also be formed at the nickel electrode as most of the systems are positive electrode (nickel) limited [25].

OER: $4\ OH^- \rightarrow O_2 + 2\ H_2O + 4\ e^-$ 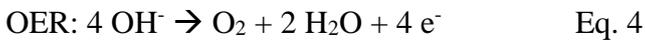 Eq. 4

HER: $2\ H_2O + 2\ e^- \rightarrow H_2 + 2\ OH^-$ 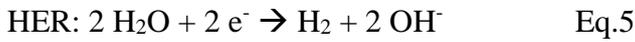 Eq.5

Besides a reduction of the coulombic efficiency, the formation of gas bubbles leads to an increase of the internal resistance by drying out the electrolyte in the electrodes porosities and in the separator. Moreover, the oxygen produced during overcharge at the nickel electrode may diffuse through the electrolyte and oxidize the metallic zinc at the negative electrode. A schematic representation of the Ni-Zn battery is shown on figure 1.

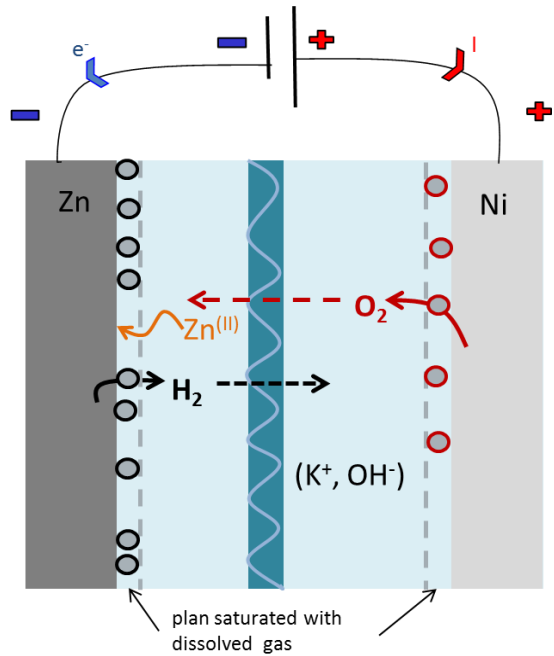

Figure 1: Schematic representation of the Ni- Zn system.

Many researches have been made in order to solve these problems [37], particularly focused on additives to the zinc electrode [38–41], improvement of the separator material [24] and composition of the electrolyte [21,28,42].

In order to study these phenomena that occur inside the battery and reduce the cycle life, we have considered the possibility of couple the two different *operando* and non-intrusive techniques of characterization described previously: strain gauges, measuring the overall deformations of the casing, and acoustic emission (AE) technique to detect transient elastic waves propagating throughout the battery subsequently to the release of internal mechanical stresses in materials.

## 2. Experimental set up

Nickel – zinc secondary batteries type AA 2500mWh 1.6V produced by Pkcell are used in this study [43]. Their rated capacity about 1,450 mAh. The recommended charge protocol is to apply a current corresponding to C/2 to C (*i.e.* 725 to 1450 mA) up to 1.9 V, followed by a constant voltage charge at 1.9V of 90 minutes, or as long as the current is higher than 50 mA. Maximum recommended discharge current is 1450 mA up to 1.2 V as cut-off voltage. The electrochemical measurements were performed using a Bio-Logic SAS potentiostat/galvanostat SP300.

The strain gauges used are rectangular rosette gauges 062LR manufactured by Micro-Measurements made for general purpose on steel [44].

The acoustic signals were recorded using a Mistras data acquisition system. To detect the acoustic waves, we used a piezoelectric sensors type nano30 provided by MISTRAS Group, Inc. [45] placed at the surface of the battery cell. The temperature range of this sensor is -65 to +177 °C, its operating frequency ranges between 125 to 750 kHz and the resonant frequency to a variation of the pression on the ceramic (V/µbar) is 300 kHz. The electric signal delivered by the piezoelectric sensor is amplified using a 2/4/6 voltage preamplifier (selectable gain ranges of 20, 40, and 60 decibels) from Mistras Group [46]. Signal acquisition and digital signal processing are obtained via a PCI-2 AE system board also manufactured by Mistras group. Using a pre-amplification gain of 60 dB and a sample rate of 2 Mega Sample Per Second (MSPS). Acoustic signals are recorded when the signal exceeded a threshold of 27 dB. The numerical datas obtained after sampling of the waveforms are imported to Matlab, in order to calculate the power spectral density (PSD) to identify different clusters of waveforms, calculating the Euclidean distance between the PSD.

The schema of the whole experimental set-up is illustrated by the Figure 2 (a), with the details of a rectangular rosette strain gauge (Figure 2 (b)).

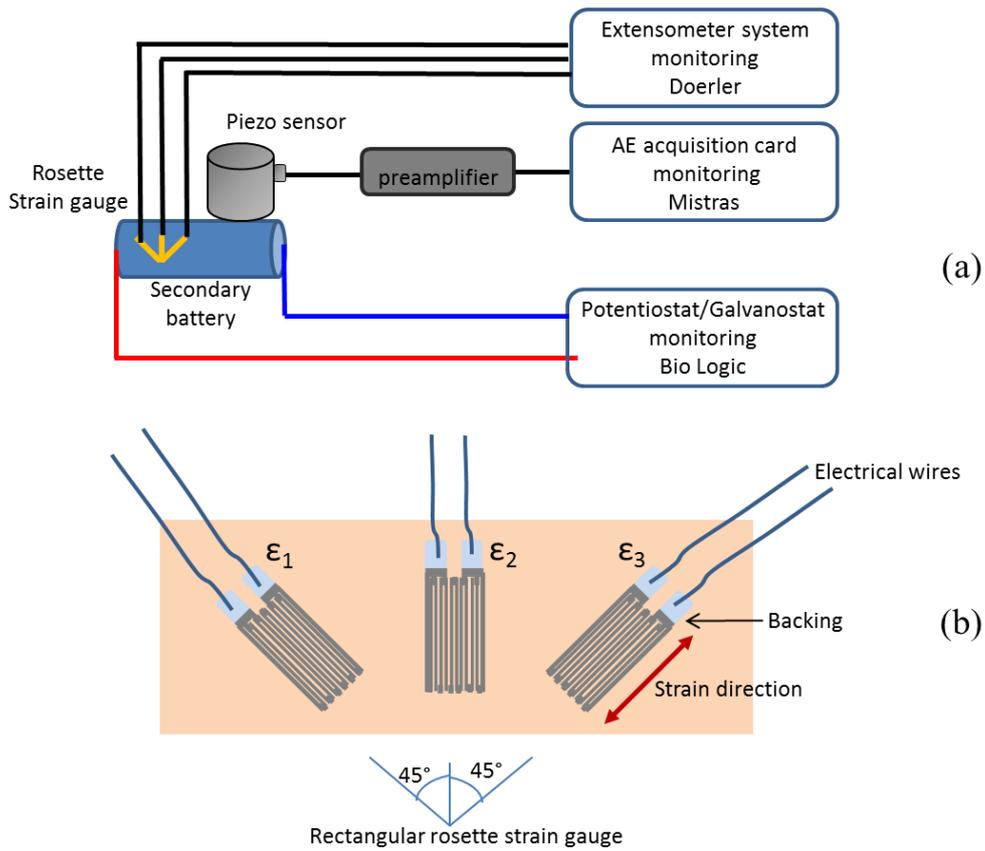

Figure 2: Experimental set-up: (a) global view, (b) rectangular rosette strain gauge.

## 3. Results and discussion

3.1 Strain gauges measurement during battery cycling

The battery was equipped with a rosette gauge, which allows recording the mechanical deformation of the structure. The gauge was placed in the middle of the cylindrical battery, with the axis 2 of the rosette gauge perpendicular to the principal axis of the cylinder. This layout allows us estimating the radial and longitudinal deformations, $E_1$ and $E_2$, calculated from the three deformation measurements $\varepsilon_1$, $\varepsilon_2$ and $\varepsilon_3$ :

$$E_1 = \sin\left(\frac{\pi}{4}\right) \times (\varepsilon_1 + \varepsilon_3) \tag{1}$$

$$E_2 = \varepsilon_2 + \cos\left(\frac{\pi}{4}\right) \times (\varepsilon_1 + \varepsilon_3) \tag{2}$$

Figure 3 presents the evolution the parameter $E_1$ (radial contribution of the deformation) and $E_2$ (longitudinal contribution of the deformation) versus the capacity during typical cycling of a Ni-Zn battery with four different cut off voltage defined during charge voltages: 1.8V, 1.85V, 1.9V and 1.95V.

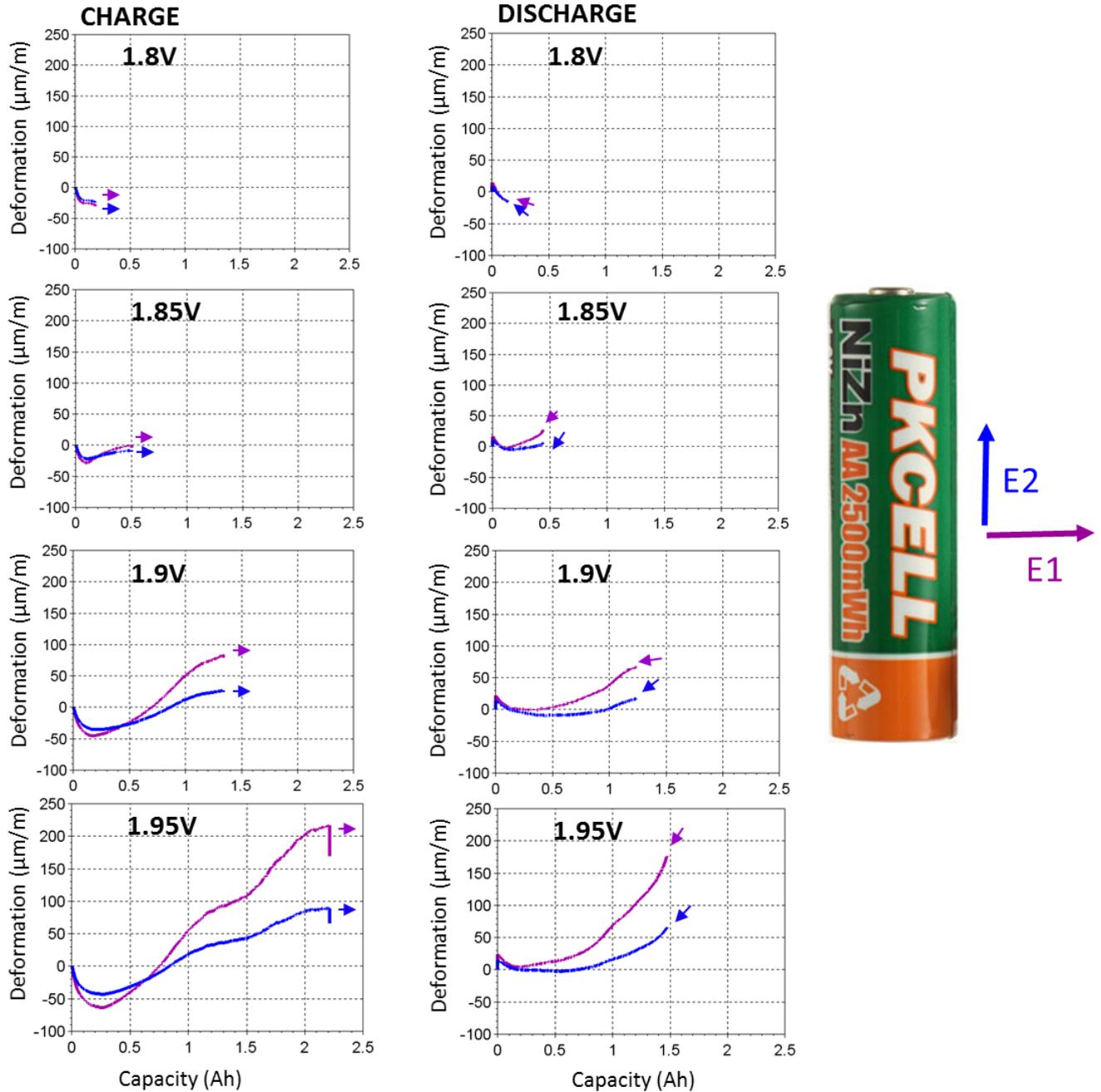

Figure 3: Evolution of the parameters E1 (purple line) and E2 (blue line) versus the capacity during typical cycling of Ni-Zn battery with four end of charge voltage: cut-off voltage of 1.8 V, 1.85 V, 1.90 V, 1.95 V.

When the cut-off voltage during charge is limited to 1.8 or 1.85 V, only small changes in both radial and longitudinal orientations are noted. In that case, the battery is only partly charged (state of charge of 14 % and 34 % at 1.8 V and 1.85 V, respectively). A small contraction of the battery (lower than 25 µm/m) is observed on both contributions at the beginning of the charge (up to 0.5 Ah). During the discharge, the deformation follows a reverse path but one can note that the deformation is slightly positive (dilatation) at the end of discharge, going back to the initial value during the rest period following the discharge.

If the cut-off voltage is set at 1.9 V, both radial and longitudinal deformations show more significant change than previously with the decrease (contraction) at the beginning of the charge, followed by an increase (dilatation) after 0.7 Ah (SoC 50 %). Even though the radial (E1) and longitudinal (E2) contributions of the deformation show similar variations, the deformation of the battery is more important in the radial direction, indicating a swelling rather than an elongation of the casing. This is commonly called a "barrel effect".

During discharge, a similar behavior as previously noted for 1.85 V cut-off voltage: a decrease of the deformation, followed by a slight increase at the tail end of the discharge. Then, the deformation goes back to its initial value during the rest period.

During a charge up to 1.95 V, the deformation continues to increase, indicating an even more significant swelling of the battery casing. It can be noted three different steps: at the beginning the deformation is negative (up to 0.2 Ah), then it increases and seems to reach a plateau at 1.45 Ah (the nominal capacity of the battery). Beyond 1.45 Ah, the battery is overcharged and the deformation rise again to a maximum value reached at a maximum capacity of 2.2 Ah. It can be noted that the swelling is partly resorbed during the rest period after the charge. During the discharge it exhibits similar behavior to that noted after a charge at 1.9 V, except it starts at a higher value of deformation.

Many phenomena could be cause of this sequence of contraction and dilatation of the casing during cycling. During charge, the contraction of the casing is probably the consequence of two mechanisms occurring at the electrodes: oxidation of $Ni(OH)_2$ to $NiOOH$ and reduction of zinc hydroxide species to $Zn^0$:

- oxidation of $Ni(OH)_2$ to $NiOOH$. Oxidation of the $-\beta-Ni(OH)_2$ to $\beta-NiOOH_2$ is obtained through the removal of a proton and leads to a contraction of the Ni-Ni distance. According to [47,48], a decrease of volume by 15% can be observed.
- zinc hydroxide species ($Zn(OH)_2$ and/or $Zn(OH)_4^{2-}$, density of $Zn(OH)_2 = 3.05$ g/cm$^3$) are converted to metallic zinc (density = 7.13 g/cm$^3$). Thus, the same amount of matter takes up less volume.

The dilatation that counterbalance the contraction can be attributed to the formation of another phase of $NiOOH$: $\gamma-NiOOH$, mainly formed during overcharge and high rates of charge. According to [47,48], a 44 % volume increase is associated to the formation of this phase. During overcharge, oxygen and hydrogen evolution take at the positive and negative electrode respectively [33].

### 3.2 Acoustic emission measurement during battery cycling

Figure 4 (a) shows a typical evolution of the acoustic activity recorded during the charge/discharge cycles of a Ni-Zn battery with two different end-of-charge voltages. The battery cycling is composed of a discharge of the residual capacity (previous charge), followed by three cycles in normal condition of charge (1.9 V cut-off) and three cycles with mild overcharge (1.95 V cut-off). The charge – discharge rate value was of C/2 (725 mA), with a discharge voltage cut-off of 1.2 V. During charge, the current applied is maintained constant (Constant current: CC) until reaching the chosen cut-off voltage 1.9 V (normal conditions) and 1.95 V (overcharge). Then the cell voltage is maintained constant (Constant Voltage: CV) until the value of the current has decreased to C/10 (140 mA). The charge phase is followed by during an open-circuit-voltage (OCV) of 0.5 hours. Each acoustic event detected is reported on the voltage vs. time curve as a black asterisk. It appears that the acoustic events are mainly detected during the end of charge (CV) and beginning of discharge. To better compare the number of acoustic events recorded during "normal cycle" (cut-off voltage of 1.9 V) and mild overcharge (cut-off voltage of 1.95 V), we reported on the Figure 4 (b) the cumulated amount of acoustic events recorded versus time. During cycling in "normal" conditions, only a very few acoustic events are recorded, mainly during discharge. When the cut-off voltage is increased to 1.95 V, we can note a significant increase of the number of acoustic events recorded, rising up to 700 events per cycle.

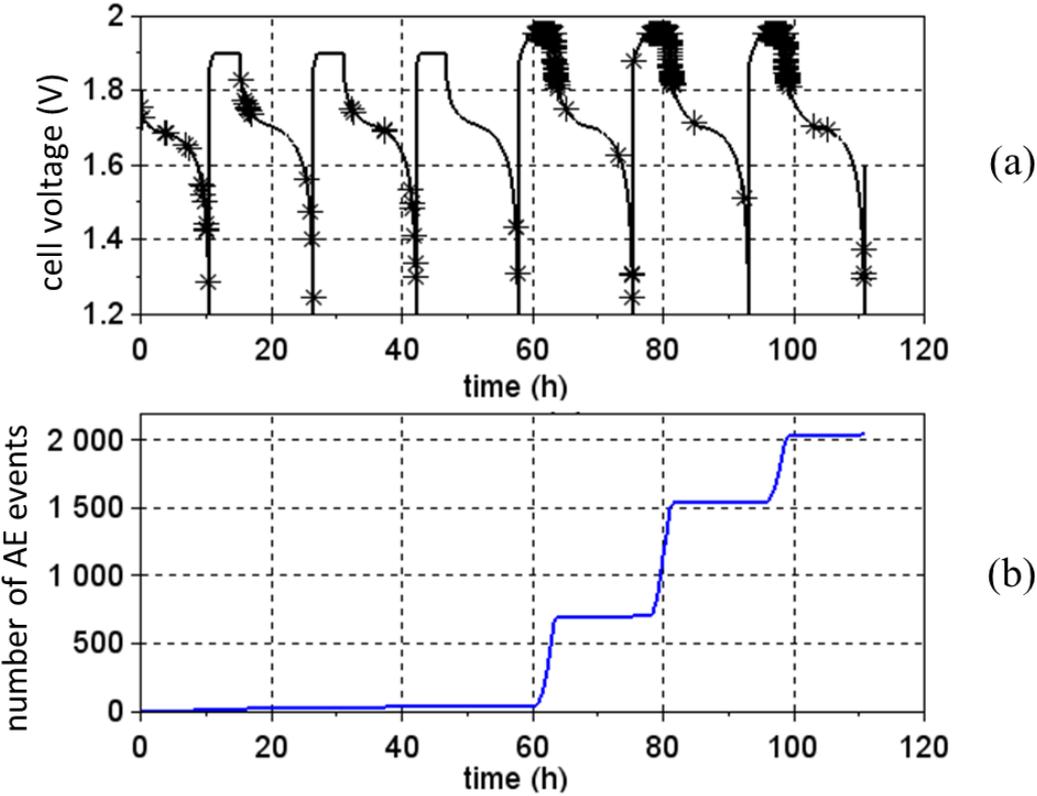

Figure 4: Evolution of acoustic emission activity during Ni-Zn battery charge/discharge cycle: (a) cell voltage curve and EA activity (*) versus time; (b) cumulated acoustic emission hits versus time.

In order to confirm the phenomenon, several batteries of the same type were cycled with different values of the cut-off voltage: 1.8 V, 1.85 V, 1.9 V, 1.95 V, and 2V. Evolution of the charge factor (defined as the number by which the quantity of electricity on discharge has to be multiplied to determine the quantity of electricity on charge required for the battery to recover its original state of charge) and the amount of acoustic events recorded during the cycle versus the charge capacity are shown on Figure 5.

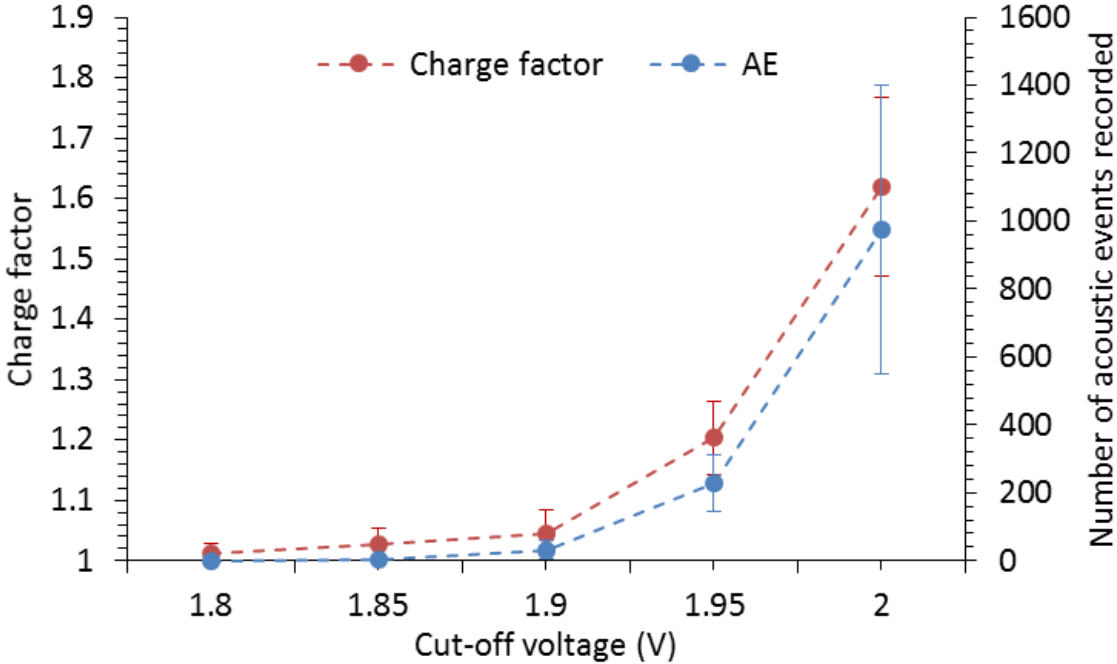

Figure 5: Evolution of the charge factor and the number acoustic events recorded during a full cycle versus the cut-off voltage.

As long as the cut-off voltage is lower than 1.9 V, the charge factor remains close to 1 (the capacity recovered during discharge is close to the capacity charged) and only few acoustic events are recorded during cycling. When the cut-off voltage is set higher than 1.9 V (i.e. 1.95 or 2 V), the discharge capacity of the battery do not exceeds 1500 mAh even though the charge capacity is much higher (e.g. close to 2500 mAh at 2 V). The value of the charge factor is 1.66. A significant part of the power supplied to the battery is consumed by parasitic reactions (heat generation and gas evolution) and can't be recovered during discharge. At the same time, the number of acoustic events increases sharply. This increase can be directly correlated to the increase of the charge factor. This result confirms the correlation between the acoustic activity of the Ni-Zn battery and its overcharge and shows a good reproducibility.

This correlation between the overcharge and the acoustic activity is not surprising. Indeed, one of the main parasitic reaction caused by an overcharge of a Ni-Zn system leads the gas evolution (oxygen evolution at the positive electrode and possible hydrogen evolution at the negative electrode). As it is schematized of the Fig 1, gas is produced at the electrolyte/electrode interface, forming a boundary layer of dissolved gas into the electrolyte. Increasing the dissolved gas concentration leads to the formation of gas bubbles when the saturation point is reached. When the bubbles pull away from the interface, they pass through in an unstable environment unsaturated in dissolved gas which drives to their collapse. Several studies report the use of acoustic emission in order to measure some characteristics of gas release as described in the review from J. W. R. Boyd and J. Varley [11]. The gas bubbles are particularly known as acoustic emission source [49] and the bubble collapse is the main phenomenon that produces transient elastic waves that can be detected as an acoustic events [10,50].

The acoustic signals registered are analyzed in order to determine if several acoustic populations could be discriminated. Each acoustic emission event is characterized by a specific waveform (as represent on Figure 6 (a)). This AE waveform could be described by different terms as for example duration, maximal amplitude, energy [15,17]… In order to describe each waveform in the frequency domain, the power spectral density (PSD) could be calculated (Figure 6 (b)) using the Welch's periodogram method [51], which is commonly used in case of non-stationary tests and allows a reduction of the noise.

Advanced data analysis technics are used so as to explore the distribution of theses PSDs in such a framework and allow exhibiting different types (clusters) of emission. Unfortunately the signals lie in a very high dimensional space (the dimension equals the sampling rate, e.g. 1024 dimension). In sake of visualization, data is displayed onto a 2D space. The chosen non-metric method [52] allowing a fruitful map is based on a nearest neighbor classifier. This kind of mapping method tends to emphasize clusters. The distribution on the resulting map (Figure 6 (c) left insert) account for neighborhood ranks between data. Each point refers to a registered waveforms; X and Y axis do not have any physical signification (the map is invariant by rotation and symmetry). Thereafter, the proximity between waveforms are coded by color of points. A unique color is associated to each waveform, as a function of its position onto the map [53]. Points with similar color correspond to neighbor waveforms according to PSD (Figure 6 (c) right insert). Consequently, different colors allows defining different clusters of signal waveforms.

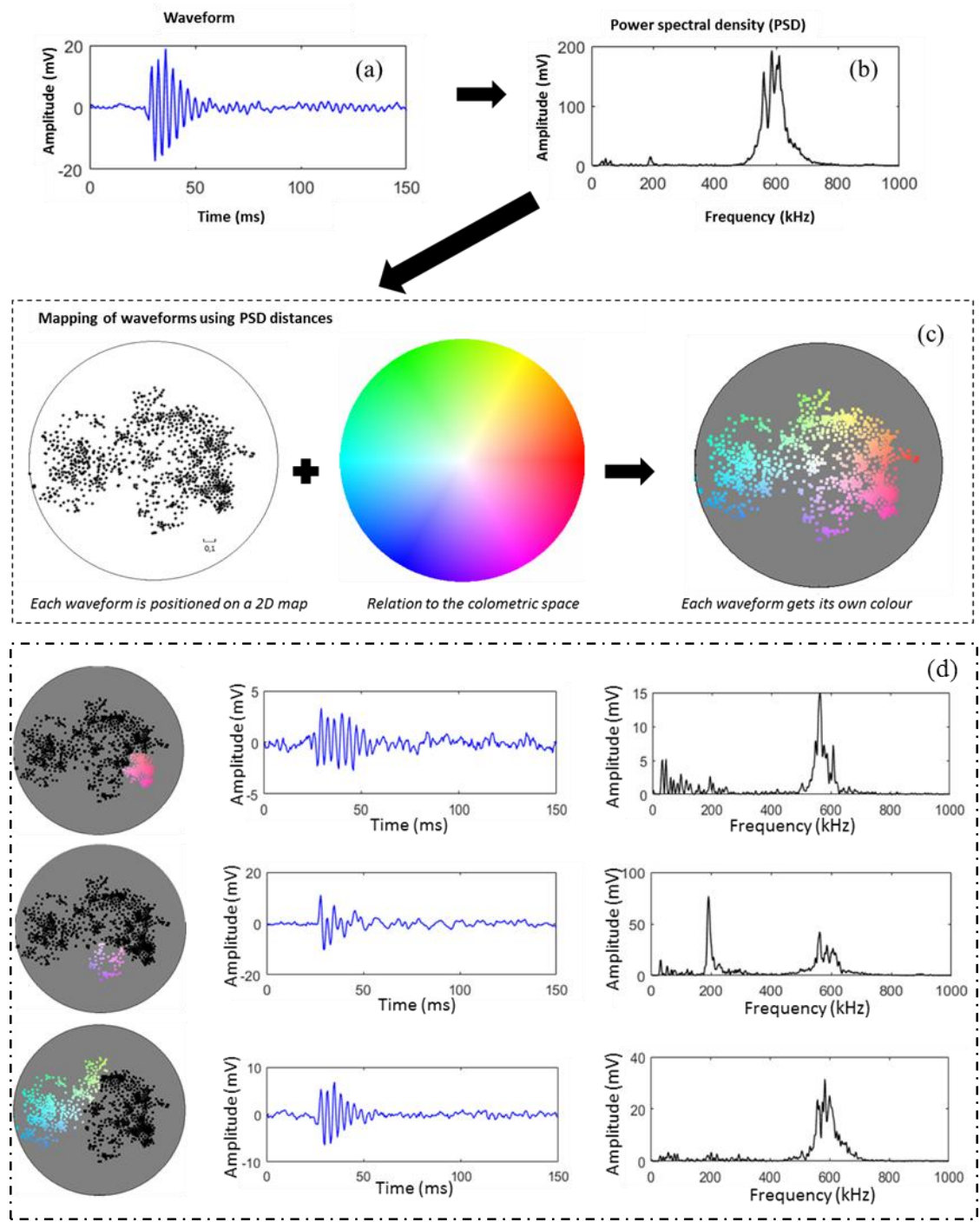

Figure 6: Methodology used to analyze the acoustic emission waveforms and identify different clusters of acoustic signals.

The displayed map on the Figure 6 (c) is obtained from the waveforms PSD recorded during the cycling test described previously, with different charge voltage cut-off: three cycles with 1.8 V, three cycles with 1.85 V, three cycles with 1.90 V and three cycles with 1.95 V.

This map allows considering at least three well discriminated clusters of waveforms, identified on the Figure 6 (d). The first cluster, localized in the pink area, is characterized by a peak frequency around 550 kHz, and a second contribution at frequencies lower than 200 kHz. The amplitude of the waveform is lower than 5 mV, thus the energy of the acoustic signal is low (< 6 aJ = 6 10$^{-18}$ J) . The second cluster, localized in the purple area, is characterized by two frequency peaks, one at 200 kHz and one slightly before 600 kHz. The amplitude of these waveforms is higher than the first cluster, varying from 10 to 20 mV. Energy of the acoustic signals of the second cluster ranges between 3 and 10 aJ. The third cluster, localized in the blue to green area, is characterized by a peak centered on 600 kHz. The amplitude of the waveforms vary from 5 to 20 mV. Thus, the energy of these acoustic signal can reach 20 aJ. The other waveforms are not very well identified, principally because they can be seen as the superposition of numerous waveforms. The focus will be kept on these three clusters throughout the rest of the study.

The interpolation of the acoustic waveforms with the voltage of the cell, following its capacity, give some more information about those waveforms. Figure 7 represents the evolution of the cell voltage following its measured capacity, for a cycle with a 1.95 V charge voltage cut-off. The battery is overcharged to 2.2 Ah, the discharged capacity is of 1.5 Ah (charge factor of 1.46).

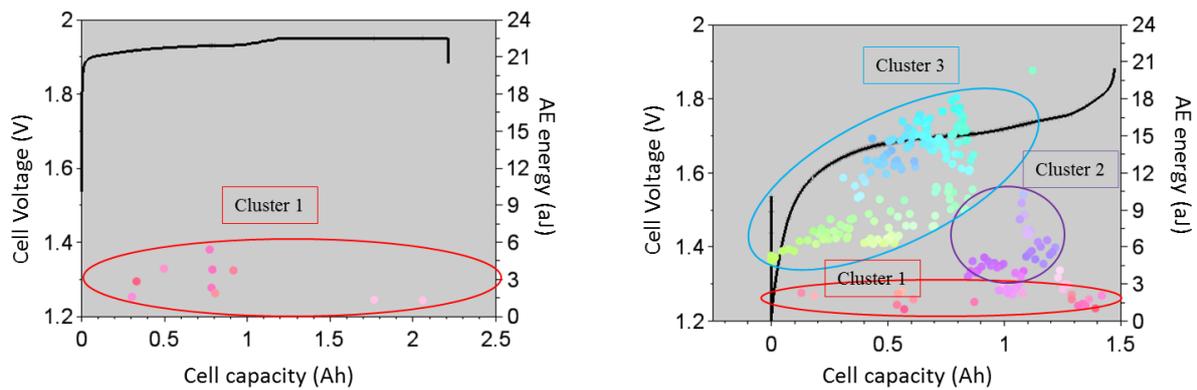

Figure 7: Evolution of cell voltage and acoustic signal energy versus capacity during Ni-Zn battery charge with a cut-off voltage of 1.95 V (left) and discharge (right).

The energy of each acoustic signals is superposed to the cell voltage curve during charge and discharge. The acoustic signals are represented by colored dots according to the position attributed by the mapping of the Figure 6 (c). The acoustic activity is principally detected during discharge. Only a few acoustic signals form the pink area (cluster 1) are observed during the charge, mainly during the first part of the charge, at constant current. At the end of the charge, during constant voltage phase at 1.95 V, almost no acoustic signal is recorded. However, as the energy of this type of signals is quite low, comprised between 0 and 6 aJ, it is possible that some signals of the same type did not exceed the acquisition threshold and thus were not been recorded. This type of acoustic signal is also observed during the discharge, they could be related to the stress released by the zinc electrode [26].

At the beginning of the discharge, when the stored capacity is still comprised between 1.5 and 0.8 Ah, acoustic signals from the purple area (cluster 2) are also observed, with an energy from 3 to 10 aJ. These acoustic signals may correspond to material cracks in the nickel electrode, due to the variation of crystallite size during overcharge [32].

From 0.9 Ah until the end of the discharge the signals from the blue and green areas (cluster 3) appear, their energy is very high at the beginning and progressively declines. These acoustic signals may be due to the collapse of gas bubbles produced by OER and HER. Indeed, the waveforms are similar to those observed in the literature [54,55]. Their time localization is coherent with the phenomenon already observe by Ito *et al.* [33]: a part of the hydrogen evolved on charging is trapped in the porosities of the negative electrode and is released during discharging.

### 3.3 Coupling acoustic emission & strain gauges measurements

As the acoustic activity was registered in the same time as the signal given by the strain gauge placed on the same battery, it is possible to couple the results thanks to a time interpolation. The figure 8 show the strain gauge and the acoustic emission activity during the charge / discharge with a cut-off voltage of 1.95 V. The electrical power (P=UI) supplied to the battery (charge) or by the battery (discharge) has been plot versus time for the charge and the discharge.

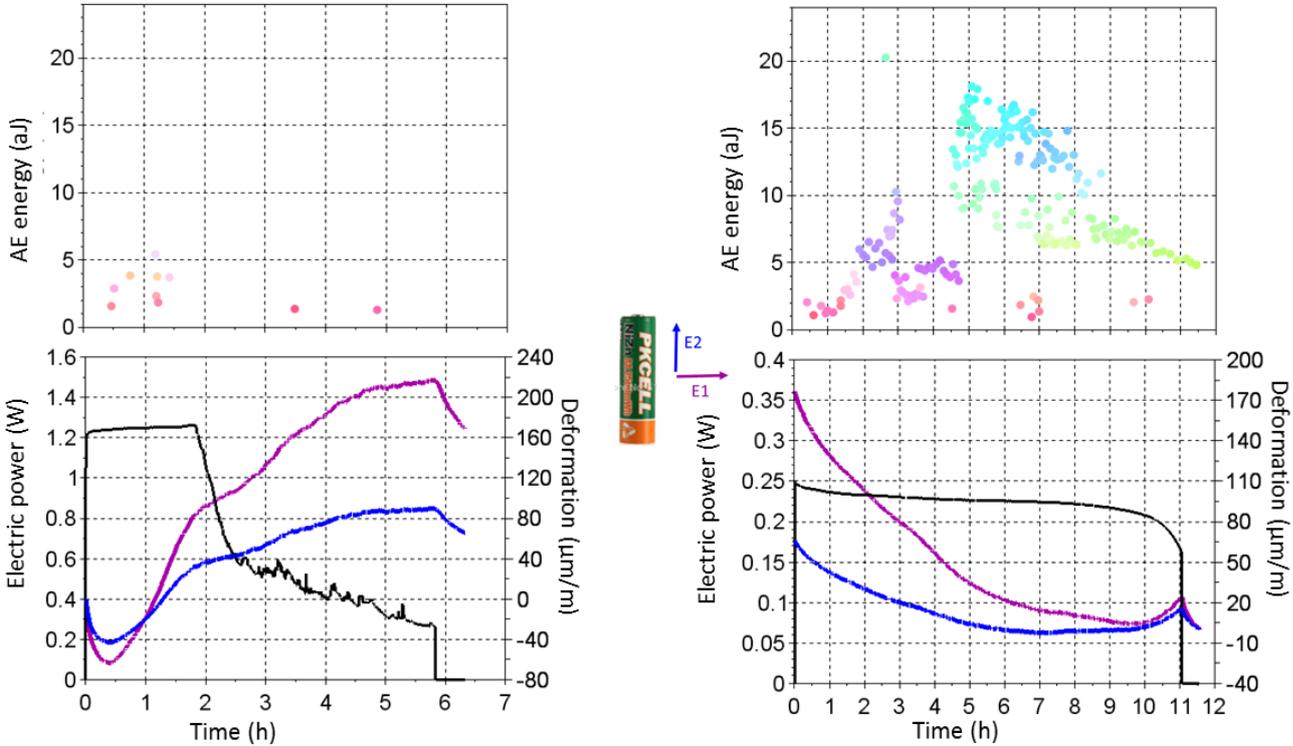

Figure 8: Evolution of acoustic emission activity and gauge strain deformation versus time during Ni-Zn battery charge/discharge for a cut-off voltage of 1.95 V.

During the overcharge period (after 2 h of charge at constant current at C/2), many fluctuations in the power can be observed. Such a behavior is typical of the consequences of the gas evolution at the surface of the electrodes: bubbles formation which modify the charge transfer and mass transport within the battery. These fluctuations are noted as the deformation curves increase again after a plateau, shortly after 2 h of charge. Thus, it is reasonable to assume that the increase of the deformation curves is related to the increase of pressure inside the battery casing, due to gas evolution on the electrodes. Interestingly, in spite of the important deformation observed, only very few acoustic events of a very low energy (cluster 1) are detected during charge. A possible explanation of this surprising low acoustic activity is that the acoustic activity related to the mechanical deformation is too weak in comparison to the threshold chosen. Another possibility is that some acoustic signals emitted are outside the range of the sensor sensitivity.

During the discharge, as the battery swelling decreases, an important acoustic activity is recorded. However, as seen before, these signals can distributed to at least three different clusters. Acoustic signals belonging to the cluster 1 are detected all along the discharge. As, they are also detected during charge, they could be attributed to the electrodes deformation.

Signals from the cluster 2 are detected at the beginning of the discharge, when the slope of the deformation curve is the greatest. A sudden change appears after 5 h (0.8 Ah remaining), as signals from the cluster 2 stop, replaced by signals from the cluster 3. At this point, the deformation curves also show a change of slope. Acoustic events of the highest energy arise as the deformation change is the slowest and continue up to the end of the experiment, even during the rest period that follows the discharge.

As mentioned above, acoustic activity related to clusters 2 and 3 could be attributed to bubbles collapse and material cracks in the electrodes but it is difficult to identify the real cause of the acoustic emission. However, as the figure 5 demonstrates, the number of acoustic emission events is highly correlated to the charge factor, that is, the amount of energy lost. We can therefore assume that acoustic signals of clusters 2 and 3 are related to gas evolution. Acoustic signals of the cluster 2, is probably due to the release of stress in the zinc electrode and signals of the cluster 3 are attributed to the collapse of gas bubbles released during dissolution of the zinc electrode. The gradual decline of the energy of acoustic signals from the cluster 3 during discharge would then be in good agreement with the continuous increase of the internal pressure of a Ni-Zn battery cell noted by Ito *et al.* [33].

This figure shows the interest of coupling two different non-intrusive techniques that give complementary information. Indeed, the strain gauge measurement could allow to optimize the charge of the battery while ensuring an acceptable swelling. Even if the sensor sensitivity and the background noise limit the acoustic emission quantification, it still provide a good indication of the overcharge of the Ni-Zn battery. Moreover, the data analysis methodology proposed allows determining different populations related to the gas evolution and the materials cracks, that offer interesting perspectives in the evaluation/optimization of batteries.

## 4. Conclusions

This work investigated an *operando* and non-invasive characterization techniques for Ni-Zn battery monitoring. Acoustic emission and strain measurement were used to monitor a commercial Ni-Zn secondary battery during cycling at several charge voltage cut-offs. Strain gauge measurement highlighted the self-breathing of Ni-Zn battery material during cycling. The deformation measurements show that a succession of contraction and dilation of the casing is noted during charge. The swelling of the casing increase significantly as the battery is charged and even more during overcharge. The mechanical deformations of the battery are accompanied by acoustic emission essentially during the discharge. If the deformation is directly correlated to the overcharge of the battery it seem that acoustic emission is mainly due to the collapse of gas bubbles evolved during overcharge. This work shows the interest of using these two complementary characterization techniques for the monitoring of battery, their management and the study of ageing mechanisms.

**Acknowledgements**

The authors thank Bpifrance for financial supporting this work via OPERA2 project.